\documentclass[10pt, final, journal, letterpaper, twocolumn]{IEEEtran}

\usepackage{amsfonts}
\usepackage[dvips]{graphicx}
\usepackage{times}
\usepackage{cite}
\usepackage{amsmath}
\usepackage{cases}
\usepackage{array}
\usepackage{dsfont}
\usepackage{amssymb}

\usepackage{stfloats}
\usepackage{slashbox}
\usepackage{graphicx}
\usepackage{footnote}
\usepackage{booktabs}
\usepackage{array}
\usepackage{bm}
\usepackage{multicol}
\usepackage{algorithmic}
\usepackage{algorithm}
\usepackage{multirow}
\usepackage{color}

\begin{document}

\title{Optimal Degrees of Freedom Region for the Asymmetric MIMO Y Channel}
\author{\IEEEauthorblockN{Kangqi Liu, Xiaojun Yuan, and Meixia Tao}\\
\thanks{Kangqi Liu and Meixia Tao are with the Department of Electronic Engineering, Shanghai Jiao Tong University, Shanghai, China (Emails: k.liu.cn@ieee.org, mxtao@sjtu.edu.cn). Xiaojun Yuan is with the School of Information Science and Technology, ShanghaiTech University, Shanghai, China (Emails: yuanxj@shanghaitech.edu.cn).}
}

\maketitle

\begin{abstract}
This letter studies the optimal degrees of freedom (DoF) region for the asymmetric three-user MIMO Y channel with antenna configuration $(M_1,M_2,M_3,N)$, where $M_i$ is the number of antennas at user $i$ and $N$ is the number of antennas at the relay node.  The converse is proved by using the cut-set theorem and the genie-message approach. To prove the achievability, we divide the DoF tuples in the outer bound into two cases. For each case, we show that the DoF tuples are achievable by collectively utilizing antenna deactivation, pairwise signal alignment and cyclic signal alignment techniques. This work not only offers a complete characterization of DoF region for the considered channel model, but also provides a new and elegant achievability proof.
\end{abstract}

\section{Introduction}
Degrees of freedom (DoF) characterizes how the capacity of a wireless channel scales in the high signal-to-noise ratio (SNR) region. The analysis of DoF and DoF region for the MIMO Y channel has attracted a lot of attention in the literature \cite{Lee1,Chaaban1,Mu1,Yuan1,Tian,Wang3,Wang8,Wang5,Liu4,Liu5,Zewail,Wang7,Chaaban4,Liu6}. The main findings are summarized in \textsc{Table} \ref{table1}. Here, $N$ denotes the number of antennas at the relay node and $M$ denotes the number of antennas at each source node. Also, $M_i$ denotes the number of antennas at source node $i$ when the source antenna setting is asymmetric. For a given channel, we say that the analysis is complete if the optimal sum DoF (or DoF region) is obtained for arbitrary values of $N$ and $M$ (or $M_i$) under the given channel category, otherwise we say it is partial. From \textsc{Table} \ref{table1}, it is seen that the existing analysis of the optimal DoF region is not complete for the $K(\geq 3)$-user MIMO Y channel under both symmetric/asymmetric antenna setting.

In this letter, we present a complete characterization of the optimal DoF region for the asymmetric three-user MIMO Y channel with antenna configuration $(M_1,M_2,M_3,N)$. While the converse of the optimal DoF region can indeed be proved easily using the cut-set theorem and genie-message approach, the achievability proof is however very challenging. The conventional method to analyze the achievable DoF in MIMO Y channel in the literature, including \cite{Lee1,Chaaban1,Mu1,Yuan1,Tian,Wang3,Wang8,Wang5,Liu4,Liu5,Zewail,Chaaban4}, is to first divide all possible antenna configurations into separate cases, and then analyze the achievable DoF case by case. If the achievable DoF coincides with the DoF outer bound, then the optimality is claimed. In our considered asymmetric antenna setting, this method, however, will be unduly complicated since the antenna configuration needs to be described by a four-dimensional tuple. In this letter, we therefore take a different and much more elegant approach to prove the DoF achievability. In specific, given the outer bound of the DoF region, we divide the DoF tuples in the outer bound into two cases. Then, for each case, we prove that the DoF tuples are achievable by using techniques including antenna deactivation, pairwise signal alignment and cyclic signal alignment. The main novelty of our proof lies in the appropriate combination of antenna deactivation, pairwise signal alignment and cyclic signal alignment to achieve the DoF outer bound without specifying the antenna configuration.

\begin{table*}[]
\tiny
\centering
\caption{Recent Advances towards the DoF Analysis for the MIMO Y Channel}\label{table1}
\begin{tabular}{|l|l|l|l|l|l|}
\hline
Channel Model                            & Antenna setting        & Sum DoF/DoF region & Antenna configuration for optimal sum DoF/DoF region                                             & Status  & Reference              \\ \hline
\multirow{3}{*}{(Three-user) MIMO Y channel}          & Symmetric                   & Sum DoF            & $\frac{N}{M} \in (0, +\infty)$                                                & Complete    & \cite{Lee1,Chaaban1} \\ \cline{2-6}
                                         & \multirow{2}{*}{Asymmetric} & Sum DoF            & $(M_1,M_2,M_3,N) \in \mathbb{R}_+^4$                                          & Complete    & \cite{Chaaban1}      \\ \cline{3-6}
                                         &                             & DoF region         &  Refer to Lemma 1-2 in \cite{Zewail}                                       &  Partial   & \cite{Zewail}          \\ \hline
\multirow{2}{*}{Four-user MIMO Y channel}          & Symmetric                   & Sum DoF            & $\frac{N}{M} \in (0, +\infty)$                                                & Complete    & \cite{Wang8,Liu4} \\ \cline{2-6}
                                         & Asymmetric & DoF region            &         Refer to Lemma 3-4 in \cite{Zewail}                                 & Partial    & \cite{Zewail}  \\ \hline
\multirow{3}{*}{$K$-user MIMO Y channel} & \multirow{2}{*}{Symmetric}  & Sum DoF            & $\frac{N}{M} \in \big(0, 2+\frac{4}{K(K-1)}\big] \cup \big[K-2, +\infty\big)$ & Partial & \cite{Liu4}          \\ \cline{3-6}
                                         &                             & DoF region         & $\frac{N}{M} \in (0, 1] \cup [K, + \infty)$                                                      & Partial & \cite{Chaaban4}      \\ \cline{2-6}
                                         & Asymmetric                  & Sum DoF            & $N \geq \max\{\sum_{i=1}^K M_i-M_s-M_t+d_{s,t}\mid \forall s,t\}$             & Partial & \cite{Liu5}          \\ \hline
$L$-cluster $K$-user MIMO multi-way relay channel             & Asymmetric                   & Sum DoF            & Refer to Theorem 2-4 in \cite{Tian}      & Partial               & \cite{Tian}               \\ \hline
(Three-user) MIMO Y channel                           & Asymmetric       & DoF region & $(M_1,M_2,M_3,N) \in \mathbb{R}_+^4$    & Complete  & This paper              \\ \hline
\end{tabular}
\vspace{-0.5cm}
\end{table*}

Notations: $(\cdot)^{T}$ and $(\cdot)^{H}$ denote the transpose and the Hermitian transpose, respectively. rank$({\bf X})$ stands for the rank of ${\bf X}$. $\textrm{span} ({\bf X})$ and ${\textrm{null} ({\bf X})}$ stand for the column space and the null space of the matrix ${\bf X}$, respectively.

\section{Channel Model}
Consider an asymmetric MIMO Y channel consisting of three users and one relay. Each user $i$ is equipped with $M_i$ antennas, for $i=1,2,3$, and the relay with $N$ antennas. Each user intends to send one independent message to each of the other two users via the relay, and there is no direct link between any two users. Denote by ${\bf H}_{i,r}(t) \in \mathbb{C}^{N \times M_i}$ the channel matrix from user $i$ to the relay for the channel use $t$, and by ${\bf H}_{r,i}(t) \in \mathbb{C}^{M_i \times N}$ the channel matrix from the relay to user $i$. It is assumed that the entries of the channel matrices are drawn independently from a continuous distribution, which guarantees that the channel matrices have full rank with probability one. Perfect channel knowledge is assumed at each node, and all the nodes in the network are assumed to be full duplex\footnote{Following the convention \cite{Lee1,Chaaban1,Mu1,Yuan1,Tian,Wang3,Wang8,Wang5,Liu4,Liu5,Zewail,Wang7,Chaaban4,Liu6}, we have assumed that the ``self-interference'' of each node can be perfectly subtracted away from its received signal, and is hence not presented in the description of channel inputs and outputs as in \eqref{y_r} and \eqref{y_i} in Section II.}. The message transmitted from user $i$ to user $j$ is denoted by $W_{i,j}$. Each $W_{i,j}$ is encoded using a codebook with size $2^{nR_{i,j}}$, where $n$ is the codeword length and $R_{i,j}$ is the information rate of $W_{i,j}$.

In the MAC phase, all the users transmit their signals to the relay. The received signal, denoted by ${\bf y}_{r}(t) \in \mathbb{C}^{N \times 1}$, at the relay is given by
\begin{eqnarray}\label{y_r}
{\bf y}_r(t)=\sum\limits_{i=1}^{3}{\bf H}_{i,r}(t){\bf x}_{i}(t)+{\bf n}_r(t),
\end{eqnarray}
where ${\bf x}_{i}(t) \in \mathbb{C}^{M_i \times 1}$ denotes transmitted signal from user $i$ and ${\bf n}_r(t) \in \mathbb{C}^{N \times 1}$ denotes the additive white Gaussian noise (AWGN) vector for the channel use $t$ with each element being independent and having zero mean and unit variance.

Upon receiving ${\bf y}_r(t)$ in \eqref{y_r}, the relay processes these messages to obtain a mixed signal ${\bf x}_{r}(t) \in \mathbb{C}^{N \times 1}$,  and broadcasts it to all the users. The received signal at user $i$ is given by
\begin{eqnarray}\label{y_i}
{\bf y}_i(t)={\bf H}_{r,i}(t){\bf x}_r(t)+{\bf n}_i(t),
\end{eqnarray}
where ${\bf n}_i(t) \in \mathbb{C}^{M_i \times 1}$ denotes the AWGN vector for the channel use $t$ with each element being independent and having zero mean and unit variance.

Each user decodes its desired messages based on the received signals and its own transmitted messages. Let $R_{i,j}(P)$ be the achievable information rate of the message $W_{i,j}$ under the power constraint $P$. We say that a rate tuple $\{R_{i,j}(P)\mid \forall i, \forall j \neq i\}$ is achievable if
\begin{equation}
\lim\limits_{n\rightarrow \infty} \textrm{Pr}\left(\hat{W}_{i,j} \neq W_{i,j}\right)=0,~~~\forall i, \forall j \neq i,
\end{equation}
where $\hat{W}_{i,j}$ is the estimate of $W_{i,j}$ at user $j$ based on the received signals and the self messages.

Then, the DoF of each message is defined as
\begin{equation}\label{dof_message}
d_{i,j} \triangleq \lim\limits_{P \rightarrow \infty} \frac{R_{i,j}(P)}{\log P}.
\end{equation}
The DoF tuple of the channel is given by
\begin{align}
{\bf d} \triangleq (d_{1,2},d_{1,3},d_{2,1},d_{2,3},d_{3,1},d_{3,2}).
\end{align}
The DoF region is defined in \eqref{dof_region} on the top of the next page,
\begin{figure*}[ht]
\begin{align}\label{dof_region}
{\cal D}=\left\{\begin{array}{c}
                  (d_{1,2},d_{1,3},d_{2,1},d_{2,3},d_{3,1},d_{3,2}) \in \mathbb{R}_{+}^6: \forall(\omega_{1,2},\omega_{1,3},\omega_{2,1},\omega_{2,3},\omega_{3,1},\omega_{3,2}) \in \mathbb{R}_{+}^6, \\
                  \sum\limits_{i=1}^3 \sum\limits_{j \neq i}  \omega_{i,j}d_{i,j} \leq \limsup_{P \rightarrow \infty}\left[\sup_{R(P)\in {\cal C}(P)} \left[\sum\limits_{i=1}^3 \sum\limits_{j \neq i}  \omega_{i,j}R_{i,j}(P)\right]\frac{1}{\log(P)}\right]
                \end{array}
\right\}
\end{align}
\hrule
\end{figure*}
where ${\cal C}(P)$ is the capacity region of the asymmetric three-user MIMO Y channel.

\section{Main Result}
\textit{Theorem 1}: For the asymmetric three-user MIMO Y channel with antenna configuration $(M_1, M_2, M_3, N)$, the optimal DoF region, denoted by ${\cal D}^{*}$, can be expressed as
\begin{subequations}\label{D}
\begin{align}\nonumber
{\cal D}^{*}=&\Big\{(d_{1,2},d_{1,3},d_{2,1},d_{2,3},d_{3,1},d_{3,2}) \in \mathbb{R}_{+}^6:\\\label{D_a}
&~~~d_{p_1,p_2}+d_{p_1,p_3} \leq M_{p_1},~\forall {\bf p} \\\label{D_b}
&~~~d_{p_2,p_1}+d_{p_3,p_1} \leq M_{p_1},~\forall {\bf p} \\\label{D_c}
&~~~d_{p_1,p_2}+d_{p_1,p_3}+d_{p_2,p_3} \leq N,~\forall {\bf p}\Big\},
\end{align}
\end{subequations}
where ${\bf p}=(p_1, p_2, p_3)$ is any permutation of $(1,2,3)$.

The DoF region above is clearly an outer bound which can be easily proved by using the cut-set theorem and the genie-message approach as in \cite{Chaaban1}. The achievability shall be proved in the next section.

\textit{Remark 1} (\textit{Comparison to \cite{Chaaban4}}): In the case with $M_1=M_2=M_3=M$ and $N\leq M$, our result reduces to Theorem 1 in \cite{Chaaban4} with $K=3$.

\textit{Remark 2} (\textit{Comparison to \cite{Zewail}}): In the case with $(M_1, M_2, M_3, N)=(3,2,2,4)$, the DoF tuple $(2,0,0,2,2,0)$ is achievable by Theorem 1. Interestingly, this DoF tuple cannot be achieved using the scheme proposed in \cite{Zewail}.

\section{Proof of DoF-Region Achievability}
We first provide some definitions on message flow graph. Note that PDE pattern have been introduced in \cite{Lee1,Chaaban1} and CDE pattern have been introduced in \cite{Wang7,Chaaban4}.

\begin{figure}[t]
\begin{centering}
\includegraphics[scale=0.9]{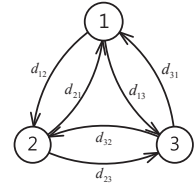}
\vspace{-0.1cm}
 \caption{Message flow graph for the three-user asymmetric MIMO Y channel.}\label{Fig_message_flow_graph}
\end{centering}
\vspace{-0.3cm}
\end{figure}

\textit{Definition 1}: Every DoF tuple ${\bf d}$ defines a \textit{message flow graph} as illustrated in Fig. \ref{Fig_message_flow_graph}, where the weight of each edge $i \rightarrow j$ represents the DoF of the message $W_{i,j}$. A message flow $i \mathop{\rightarrow}\limits^{d} j \mathop{\rightarrow}\limits^{d} i$ is called a PDE pattern $i \rightarrow j \rightarrow i$ with weight $d$. A message flow $i \mathop{\rightarrow}\limits^{d} j \mathop{\rightarrow}\limits^{d} k \mathop{\rightarrow}\limits^{d} i$ is called a CDE pattern $i \rightarrow j \rightarrow k \rightarrow i$ with weight $d$.

\textit{Remark 3}: A directed edge with weight $d_1+d_2$ can be split into two directed edges with weight $d_1$ and $d_2$.

To prove the achievability, we divide the DoF tuples in the DoF-region outer bound ${\cal D}^{*}$ into two cases: (I) $d_{1,2} \geq d_{2,1}$, $d_{3,1} \geq d_{1,3}$ and $d_{2,3} \geq d_{3,2}$, (II) $d_{1,2} \geq d_{2,1}$, $d_{1,3} > d_{3,1}$ and $d_{2,3} \geq d_{3,2}$, and present the achievable scheme for each case. It is worth mentioning that every DoF tuple in the outer bound can be converted into one of these two cases by user index-reordering. For example, when $d_{1,2} < d_{2,1}$, $d_{3,1} < d_{1,3}$ and $d_{2,3} < d_{3,2}$, we can exchange the order of user 2 and user 3, i.e., user $2'$ as original user 3 and user $3'$ as original user 2, so as to convert it to case I.

\subsection{Case I}
In this subsection, we explain how to achieve the DoF tuple ${\bf d} \in {\cal D}^{*}$ satisfying $d_{1,2} \geq d_{2,1}$, $d_{3,1} \geq d_{1,3}$ and $d_{2,3} \geq d_{3,2}$. Given the DoF tuple, we form the PDE pattern $1 \rightarrow 2 \rightarrow 1$ with weight $d_{2,1}$, $1 \rightarrow 3 \rightarrow 1$ with weight $d_{1,3}$, and $2 \rightarrow 3 \rightarrow 2$ with weight $d_{3,2}$. From the remaining data streams, we form the CDE pattern $1 \rightarrow 2 \rightarrow 3 \rightarrow 1$ with weight $\gamma$, where $\gamma=\min\{d_{1,2}-d_{2,1},d_{2,3}-d_{3,2},d_{3,1}-d_{1,3}\}$. Without loss of generality, we assume $\gamma=d_{1,2}-d_{2,1}$. Then, user $2$ has $d_{2,3}-d_{3,2}-\gamma$ additional data streams to transmit to user $3$, and user $3$ has $d_{3,1}-d_{1,3}-\gamma$ additional data streams to transmit to user $1$.

We assume that the relay only utilizes $J=d_{2,1}+d_{2,3}+d_{3,1}$ antennas for this case by the relay antenna deactivation\footnote{For any rational number $d_{i,j}$, we can use $t$-symbol extension such that $td_{i,j}$ is an integer. We refer interested readers to \cite{Liu4,Wang3} for details.}. Note that we have $J \leq N$ from \eqref{D_c}. In the $J$-dimensional subspace of the relay, the first $d_{2,1}$ dimensions are used for the PDE pattern $1 \rightarrow 2 \rightarrow 1$ with weight $d_{2,1}$ to form network-coded symbols. Similarly, the second $d_{1,3}$ dimensions are used for the PDE pattern $1 \rightarrow 3 \rightarrow 1$ with weight $d_{1,3}$, and the third $d_{3,2}$ dimensions are used for the PDE pattern $2 \rightarrow 3 \rightarrow 2$ with weight $d_{3,2}$. The fourth $2\gamma$ dimensions are used for the CDE pattern $1 \rightarrow 2 \rightarrow 3 \rightarrow 1$ with weight $\gamma$. The remaining $(d_{2,3}-d_{3,2}-\gamma)$ and $(d_{3,1}-d_{1,3}-\gamma)$ dimensions are used to decode the additional $(d_{2,3}-d_{3,2}-\gamma)$ data streams sent from user $2$ to user $3$ and the additional $(d_{3,1}-d_{1,3}-\gamma)$ data streams sent from user $3$ to user $1$, respectively.

\textit{Remark 4}: By comparing with complete decoding of all the data streams at the relay, each PDE pattern with weight $d$ reduces $d$ dimension required at the relay, and each CDE pattern with weight $d$ also reduces $d$ dimension required at the relay.

We now present the signal alignment design to realize the above patterns. We first consider the MAC phase. During the MAC phase, the signal received at the relay can be rewritten as
\begin{align}\nonumber
{\bf y}_r=&\sum\limits_{i=1}^3 \sum\limits_{j \neq i} {\bf H}_{i,r}{\bf V}_{i,j}^p{\bf s}_{i,j}^p+\sum\limits_{(i,j) \in {\cal S}} {\bf H}_{i,r}{\bf V}_{i,j}^c{\bf s}_{i,j}^c\\\label{y_r_I}
&+{\bf H}_{2,r}{\bf V}_{2,3}^r{\bf s}_{2,3}^r+{\bf H}_{3,r}{\bf V}_{3,1}^r{\bf s}_{3,1}^r+{\bf n}_r,
\end{align}
where ${\cal S}=\{(1,2),(2,3),(3,1)\}$. Here, ${\bf s}_{2,3}^p \in {\mathbb C}^{d_{3,2} \times 1}$ denotes the signal transmitted from user $2$ to user $3$ in the PDE pattern $2 \rightarrow 3 \rightarrow 2$ with weight $d_{3,2}$, ${\bf s}_{2,3}^c \in {\mathbb C}^{\gamma \times 1}$ denotes the signal transmitted from user $2$ to user $3$ in the CDE pattern $1 \rightarrow 2 \rightarrow 3 \rightarrow 1$ with weight $\gamma$, and ${\bf s}_{2,3}^r \in {\mathbb C}^{(d_{2,3}-d_{3,2}-\gamma) \times 1}$ denotes the additional signal transmitted from user $2$ to user $3$. ${\bf V}_{2,3}^p \in {\mathbb C}^{M_2 \times d_{3,2}}$, ${\bf V}_{2,3}^c \in {\mathbb C}^{M_2 \times \gamma}$, and ${\bf V}_{2,3}^r \in {\mathbb C}^{M_2 \times (d_{2,3}-d_{3,2}-\gamma)}$ are the precoding matrices for ${\bf s}_{2,3}^p$, ${\bf s}_{2,3}^c$, and ${\bf s}_{2,3}^r$, respectively. Similar definitions apply to the other $(i,j)$.

From the aforementioned relay space division scheme, we aim to design all the precoding matrices jointly such that
\begin{subequations}\label{GSA_E}
\begin{align}\label{GSA_E_1}
&{\bf H}_{1,r}{\bf V}_{1,2}^p={\bf H}_{2,r}{\bf V}_{2,1}^p\triangleq {\bf B}_1,\\\label{GSA_E_2}
&{\bf H}_{1,r}{\bf V}_{1,3}^p={\bf H}_{3,r}{\bf V}_{3,1}^p\triangleq {\bf B}_2,\\\label{GSA_E_3}
&{\bf H}_{2,r}{\bf V}_{2,3}^p={\bf H}_{3,r}{\bf V}_{3,2}^p\triangleq {\bf B}_3,\\\label{GSA_E_4}
&{\bf H}_{1,r}{\bf V}_{1,2}^{c}+{\bf H}_{2,r}{\bf V}_{2,3}^{c}={\bf H}_{3,r}{\bf V}_{3,1}^{c},
\end{align}
\begin{align}\label{GSA_E_5}
&\textrm{rank}([{\bf V}_{2,3}^p~{\bf V}_{2,3}^c~{\bf V}_{2,3}^r])=d_{2,3},\\\label{GSA_E_6}
&\textrm{rank}([{\bf V}_{3,1}^p~{\bf V}_{3,1}^c~{\bf V}_{3,1}^r])=d_{3,1}.
\end{align}
\end{subequations}
Here, condition \eqref{GSA_E_1} requires that the relay aligns the signal pair $({\bf s}_{1,2}^p, {\bf s}_{2,1}^p)$ in a subspace to form network-coded symbols; condition \eqref{GSA_E_2} and \eqref{GSA_E_3} require to align the signal pair $({\bf s}_{1,3}^p, {\bf s}_{3,1}^p)$ and  $({\bf s}_{2,3}^p, {\bf s}_{3,2}^p)$, respectively. The condition \eqref{GSA_E_4} requires that the relay aligns the signal ${\bf s}_{3,1}^{c}$ to the subspace spanned by $({\bf s}_{1,2}^c, {\bf s}_{2,3}^c)$ to form network-coded symbols. Condition \eqref{GSA_E_5} is to ensure the separability of ${\bf s}_{2,3}^p$, ${\bf s}_{2,3}^c$, and ${\bf s}_{2,3}^r$ and at user 2, and likewise condition \eqref{GSA_E_6} is to ensure the separability of ${\bf s}_{3,1}^p$, ${\bf s}_{3,1}^c$, and ${\bf s}_{3,1}^r$ at user 3.

We rewrite \eqref{GSA_E_1}-\eqref{GSA_E_3} as
\begin{equation}\label{GSA_EE_1}
\left[{{\bf V}_{i,j}^p}^T~~{{\bf V}_{j,i}^p}^T\right]^T \subseteq \textrm{null}\left(\left[{\bf H}_{i,r}~~-{\bf H}_{j,r}\right]\right).
\end{equation}
Then, we have the following lemma.

\textit{Lemma 1} \textit{(Sufficient condition for \eqref{GSA_E_1}-\eqref{GSA_E_3})}: There exist ${\bf V}_{i,j}^p$ and ${\bf V}_{j,i}^p$ satisfying \eqref{GSA_E_1}-\eqref{GSA_E_3} with probability one if
\begin{equation}\label{GSA}
M_i+M_j-J \geq \min\{d_{i,j},d_{j,i}\}, \forall i, \forall j \neq i.
\end{equation}
\begin{proof}
The proof follows directly from the rank-nullity theorem and the channel randomness. We refer interested readers to \cite{Lee1} for details.
\end{proof}

We rewrite \eqref{GSA_E_4} as
\begin{equation}\label{CSA2_b}
\left[{{\bf V}_{1,2}^c}^T~~{{\bf V}_{2,3}^c}^T~~{{\bf V}_{3,1}^c}^T\right]^T \subseteq \textrm{null}\left(\left[{\bf H}_{1,r}~{\bf H}_{2,r}~-{\bf H}_{3,r}\right]\right).
\end{equation}
Then, we have the following lemma.

\textit{Lemma 2} \textit{(Sufficient condition for \eqref{GSA_E_4})}: There exist ${\bf V}_{1,2}^c$, ${\bf V}_{2,3}^c$ and ${\bf V}_{3,1}^c$ satisfying \eqref{GSA_E_4} with probability one if
\begin{equation}\label{CSA}
M_1+M_2+M_3-J \geq d_{1,2}-d_{2,1}.
\end{equation}
\begin{proof}
Note that the column rank of $\left[{\bf H}_{1,r}~{\bf H}_{2,r}~-{\bf H}_{3,r}\right]$ is less than or equal to $J$. This implies that the dimension of the null space of $\left[{\bf H}_{1,r}~{\bf H}_{2,r}~-{\bf H}_{3,r}\right]$ is greater than or equal to $M_1+M_2+M_3-J$ with probability one. Hence, we can find at least $d_{1,2}-d_{2,1}$ linear independent vectors in the null space of $\left[{\bf H}_{1,r}~{\bf H}_{2,r}~-{\bf H}_{3,r}\right]$ if \eqref{CSA} holds.
\end{proof}

From \eqref{D_a} and \eqref{D_b}, we obtain (i) $(M_1+M_2)-J \geq d_{2,1}$; (ii) $(M_1+M_3)-J \geq d_{1,3}$; (iii) $(M_2+M_3)-J \geq d_{3,2}$; (iv) $(M_1+M_2+M_3)-J \geq d_{1,2}-d_{2,1}$. Thus, from \textit{Lemma 1} and \textit{Lemma 2}, we design the precoding matrices $\{{\bf V}_{i,j}^p \mid i\neq j\}$ and $\{{\bf V}_{1,2}^{c},{\bf V}_{2,3}^{c},{\bf V}_{3,1}^{c}\}$ such that \eqref{GSA_E_1}-\eqref{GSA_E_4} hold. The remaining two precoding matrices ${\bf V}_{2,3}^r$ and ${\bf V}_{3,1}^r$ can be designed randomly provided that \eqref{GSA_E_5} and \eqref{GSA_E_6} hold.

The signal received at the relay can be expressed as
\begin{align}\nonumber
{\bf y}_r=&{\bf B}_1({\bf s}_{1,2}^p+{\bf s}_{2,1}^p)+{\bf B}_2({\bf s}_{1,3}^p+{\bf s}_{3,1}^p)+{\bf B}_3({\bf s}_{2,3}^p+{\bf s}_{3,2}^p)\\\nonumber
&+{\bf B}_4({\bf s}_{1,2}^c+{\bf s}_{3,1}^c)+{\bf B}_5({\bf s}_{2,3}^c+{\bf s}_{3,1}^c)\\
&+{\bf B}_6{\bf s}_{2,3}^r+{\bf B}_7{\bf s}_{3,1}^r+{\bf n}_r,
\end{align}
where ${\bf B}_4 \triangleq {\bf H}_{1,r}{\bf V}_{1,2}^{c}$, ${\bf B}_5 \triangleq {\bf H}_{2,r}{\bf V}_{2,3}^{c}$, ${\bf B}_6 \triangleq {\bf H}_{2,r}{\bf V}_{2,3}^{r}$ and ${\bf B}_7 \triangleq {\bf H}_{3,r}{\bf V}_{3,1}^{r}$. Thus far, the relay is able to decode the network-coded symbols, $\{{\bf s}_{1,2}^p+{\bf s}_{2,1}^p,{\bf s}_{1,3}^p+{\bf s}_{3,1}^p,{\bf s}_{2,3}^p+{\bf s}_{3,2}^p,{\bf s}_{1,2}^c+{\bf s}_{3,1}^c,{\bf s}_{2,3}^c+{\bf s}_{3,1}^c\}$ together with the remaining symbols, $\{{\bf s}_{2,3}^r,{\bf s}_{3,1}^r\}$, by using a $J \times J$ zero-forcing matrix
\begin{align}\label{W}
{\bf W}=\left([{\bf B}_1~{\bf B}_2~{\bf B}_3~{\bf B}_4~{\bf B}_5~{\bf B}_6~{\bf B}_7]\right)^{-1}.
\end{align}
Here, since all the channel coefficients are independently drawn from continuous distributions, the above matrix has full rank almost surely.

We next introduce the transmission scheme for the BC phase. The signal received at user $i$ with receiving matrix ${\bf U}_{i} \in {\mathbb C}^{\sum_{j \neq i} d_{j,i} \times M_i}$ can be expressed as
\begin{align}\label{s_i_E}
\hat{{\bf s}}_i=&{\bf U}_i{\bf H}_{r,i}{\bf T}{\bf s}_r+{\bf U}_i{\bf H}_{r,i}{\bf T}{\bf W}{\bf n}_r+{\bf U}_i{\bf n}_i,
\end{align}
where ${\bf T} \in {\mathbb C}^{N \times J}$ denotes a zero-forcing matrix in the BC phase.

Due to the symmetry between the MAC and BC phases, we partition ${\bf U}_1$ as
\begin{align}
&{\bf U}_1=\left[{{\bf U}_{2,1}^p}^T~~{{\bf U}_{3,1}^p}^T~~{{\bf U}_{3,1}^c}^T~~{{\bf U}_{3,1}^r}^T\right]^T,
\end{align}
where ${\bf U}_{2,1}^p \in {\mathbb C}^{d_{2,1} \times M_1}$, ${\bf U}_{3,1}^p \in {\mathbb C}^{d_{1,3} \times M_1}$, ${\bf U}_{3,1}^c \in {\mathbb C}^{\gamma \times M_1}$, and ${\bf U}_{3,1}^r \in {\mathbb C}^{(d_{3,1}-d_{1,3}-\gamma) \times M_1}$. ${\bf U}_2$ and ${\bf U}_3$ are partitioned similarly. Then, we design $\{{\bf U}_i\}_{i=1}^3$ such that
\begin{subequations}\label{GSA_EEE}
\begin{align}\label{GSA_EEE_1}
&{\bf U}_{2,1}^p{\bf H}_{r,1}={\bf U}_{1,2}^p{\bf H}_{r,2},\\\label{GSA_EEE_2}
&{\bf U}_{3,1}^p{\bf H}_{r,1}={\bf U}_{1,3}^p{\bf H}_{r,3},\\\label{GSA_EEE_3}
&{\bf U}_{3,2}^p{\bf H}_{r,2}={\bf U}_{2,3}^p{\bf H}_{r,3},\\\label{GSA_EEE_4}
&{\bf U}_{3,1}^{c}{\bf H}_{r,1}+{\bf U}_{1,2}^c{\bf H}_{r,2}={\bf U}_{2,3}^{c}{\bf H}_{r,3}.
\end{align}
\end{subequations}
Comparing \eqref{GSA_EEE_1}-\eqref{GSA_EEE_4} with \eqref{GSA_E_1}-\eqref{GSA_E_4}, we see the symmetry between the design of $\{{\bf V}_{i,j}^p,{\bf V}_{i,j}^c,{\bf V}_{i,j}^r\}$ and that of $\{{\bf U}_{i,j}^p,{\bf U}_{i,j}^c,{\bf U}_{i,j}^r\}$. Then, the zero-forcing matrix ${\bf T}$ in the BC phase can be designed as
\begin{align}\label{T}
{\bf T}=\left(\left[\begin{array}{c}
                      {\bf U}_{2,1}^p{\bf H}_{r,1} \\
                      {\bf U}_{3,1}^p{\bf H}_{r,1} \\
                      {\bf U}_{3,2}^p{\bf H}_{r,2} \\
                      {\bf U}_{3,1}^{c}{\bf H}_{r,1} \\
                      {\bf U}_{2,3}^{c}{\bf H}_{r,3}\\
                      {\bf U}_{2,3}^{r}{\bf H}_{r,3}\\
                      {\bf U}_{3,1}^{r}{\bf H}_{r,1}
                    \end{array}
\right]\right)^{-1}.
\end{align}
The signal received at user $i$ in \eqref{s_i_E} can be rewritten as
\begin{align}\label{s_1_decode}
&\hat{{\bf s}}_1=\left[
                                                                                                      \begin{array}{c}
                                                                                                        {\bf s}_{1,2}^p+{\bf s}_{2,1}^p \\
                                                                                                        {\bf s}_{1,3}^p+{\bf s}_{3,1}^p \\
                                                                                                        {\bf s}_{1,2}^c+{\bf s}_{3,1}^c \\
                                                                                                        {\bf s}_{3,1}^r \\
                                                                                                      \end{array}
                                                                                                    \right]
+{\bf U}_1{\bf H}_{r,1}{\bf T}{\bf W}{\bf n}_r+{\bf U}_1{\bf n}_1
\end{align}
\begin{align}\label{s_2_decode}
&\hat{{\bf s}}_2=\left[
                                                                                                      \begin{array}{c}
                                                                                                        {\bf s}_{1,2}^p+{\bf s}_{2,1}^p \\
                                                                                                        {\bf s}_{2,3}^p+{\bf s}_{3,2}^p \\
                                                                                                        {\bf s}_{2,3}^c-{\bf s}_{1,2}^c
                                                                                                      \end{array}
\right]
+{\bf U}_2{\bf H}_{r,2}{\bf T}{\bf W}{\bf n}_r+{\bf U}_2{\bf n}_2
\end{align}
\begin{align}\label{s_3_decode}
&\hat{{\bf s}}_3=\left[
                                                                                                      \begin{array}{c}
                                                                                                        {\bf s}_{1,3}^p+{\bf s}_{3,1}^p \\
                                                                                                        {\bf s}_{2,3}^p+{\bf s}_{3,2}^p \\
                                                                                                        {\bf s}_{2,3}^c+{\bf s}_{3,1}^c \\
                                                                                                        {\bf s}_{2,3}^r \\
                                                                                                      \end{array}
                                                                                                    \right]
+{\bf U}_3{\bf H}_{r,3}{\bf T}{\bf W}{\bf n}_r+{\bf U}_3{\bf n}_3.
\end{align}
Finally, each user decodes its desired signal after self-interference cancellation. The DoF tuple ${\bf d} \in {\cal D}^{*}$ satisfying $d_{1,2} \geq d_{2,1}$, $d_{2,3} \geq d_{3,2}$ and $d_{3,1} \geq d_{1,3}$ is thus achievable.

\textit{Remark 5}: The design of the zero-forcing matrix in the MAC and BC phase is not symmetric.

\subsection{Case II}
For this case, we only consider the PDE pattern to achieve the DoF tuple ${\bf d} \in {\cal D}^{*}$ satisfying $d_{1,2} \geq d_{2,1}$, $d_{1,3} > d_{3,1}$ and $d_{2,3} \geq d_{3,2}$ by using pairwise signal alignment and antenna deactivation techniques. Let the relay only utilize $J=d_{1,2}+d_{1,3}+d_{2,3}$ antennas. Then, the method is similar to Case I and thus omitted here.

\section{Conclusion}
In this letter, we have studied the optimal DoF region of the asymmetric MIMO Y channel. The converse is proved by using the cut-set theorem and the genie-message approach. To prove the achievability, the DoF tuples in the outer bound is divided into two cases. For each case, the DoF tuples are shown to be achievable by collectively utilizing antenna deactivation, pairwise signal alignment and cyclic signal alignment techniques. In the future work, it is interesting to extend this proof method to the analysis of the more complicated relay networks.

\bibliographystyle{IEEEtran}
\bibliography{IEEEabrv,reference}

\end{document}